\documentclass[preprint,12pt]{elsarticle}

\usepackage{epsfig}

\usepackage{amssymb}

\journal{Journal of Crystal Growth}

\begin{document}

\begin{frontmatter}



\title{Hybrid plasmonic-photonic crystal formed on gel-immobilized colloidal crystal via solvent substitution}


\author[1]{Sho Kawakami}

\author[2]{Atsushi Mori\footnote{Corresponding author; e-mail atsushimori@tokushima-u.ac.jp}}

\author[3]{Ken Nagashima}

\author[2]{Masanobu Haraguchi}

\author[2]{Toshihiro Okamoto}

\address[1]{Department of Optical System Engineering, Tokushima University, Tokushima 770-8506, Japan}

\address[2]{Institute of Science and Technology, Tokushima University, Tokushima 770-8506, Japan}

\address[3]{Institute of Low Temperature Science, Hokkaido University, Kita-19, Nishi-8, Kita-ku, Sapporo, Hokkaido 060-0819, Japan}

\begin{abstract}
Gel-immobilized colloidal crystals were prepared to obtain hybrid plasmonic-photonic crystals, in which electric field enhancement to a greater extent than that due to localized surface plasmons (LSP) alone was expected due to coupling between  LSP and the photonic band.
Polystyrene colloidal crystals immobilized by the $N$-(hydroxymethyl)acrylamide gel were immersed in an aqueous dispersion of gold nanoparticles (AuNPs).
Then, the gel-immobilized colloidal crystals were picked out and immersed in an ionic liquid mixture.
The surfaces of the gel-immobilized colloidal crystals immersed in the AuNP dispersion were observed via scanning electron microscopy after this solvent substitution.
The lattice spacing of the colloidal crystal varied as the composition of the ionic liquid mixture was changed.
The composition was determined so that the photonic band gap wavelength coincided with the LSP wavelength.
Further, the reflection spectra were measured.
Thus, we successfully prepared a hybrid plasmonic-photonic crystal. 
\end{abstract}


\end{frontmatter}


\section{Introduction}
\label{intro}
Plasmonics is an exciting branch of nanophotonics.
Electric field enhancement can be realized by plasmonic resonance.
Plasmonic resonance is due to coupling between an external field and induced charge density oscillation of free electrons of noble metals.
In particular, localized surface plasmons (LSP), which occur on nanostructured surfaces such as on a metal nanoparticle, result in strong field enhancement, as described in textbooks such as Refs.~\cite{Sarid2010,Pelton2013}.
This phenomenon has already been applied in chemical sensing such as via surface-enhanced Raman spectroscopy \cite{Muehlethaler2016,Hao2015,Schlicker2014,Alvarez-Puebla2012ACIE,Sharma2012,Alvarez-Puebla2012CSR}.
Several studies on the application of plasmonics for manipulation of weak light are also in progress.

Colloidal crystals have been attracting attention in relation to crystal growth in recent decades.
Along with interest in elucidation of \textit{in situ} observation of the growth mechanism (\textit{e.g.}, Ref.~\cite{Zhang2014}), colloidal crystals, as photonic crystals, draw attention as functional materials (see, an introductory part of textbooks of phonic crystals such as Ref.~\cite{Sakoda2005}).
The photonic band structure occurs because the periodicity of special modulation of the dielectric constant is on the order of the optical wavelength.
In order to utilize the colloidal crystals as photonic crystals, the number and extent of defects in crystals should be reduced.
Various efforts such as Refs~\cite{Blaaderen1997,Zhu1997,Sawada2001,Mori2011,Hilhorost2013,Mori2014} have been made to realize this.

Coupling plasmonics and photonic crystals is a new field of hybrid nanophotonics.
A plasmonic nanostructure formed on a photonic crystal is referred to as a hybrid plasmonic-photonic crystal. Electric field enhancement to a greater extent than that due to the LSP resonance alone was expected through this structure and confirmed, \textit{e.g.}, in Refs.~\cite{Kim2010,Tao2011,Ding2013,Robbiano2013,Xu2013,Belardini2014,Zhou2014,Shao2015,Yin2016}.
In such studies, a fixed photonic crystal such as a dried colloidal crystal was used. Hence, the photonic band structure was fixed.
The design of the photonic band structure to match it to the LSP resonance should be done prior to the fabrication.
However, the sizes of the prepared metal nanoparticles and colloidal spheres deviate from the designed values, and the lattice parameters of the colloidal crystals cannot be equal to the ideal values.
One cannot, however, tune the photonic band structure after fabrication in the case of dried colloidal crystals. It was surprising, in this respect, that the electric field enhancement by a factor of several hundreds was accomplished despite this difficulty, e.g., in Refs.~\cite{Tao2011,Shao2015}. 
To overcome this limitation, i.e., the inability to tune the photonic band structure after fabrication, we have proposed to use a gel-immobilized colloidal crystal as a photonic crystal \cite{Kawakami2016}.
The photonic band structure of the gel-immobilized colloidal crystals is tunable by external stimuli \cite{Iwayama2003}.
Chemical tuning is also possible \cite{Toyotama2005,Kanai2011,Yamamoto2012}.
That is, precise tuning can be performed after measuring the photonic properties of the prepared samples.
The extent of enhancement increases with increasing degree of coincidence between the resonant wavelengths.

We have already presented preliminary results in a previous report \cite{Kawakami2016}.
Surfaces of samples after drying have been observed by atomic force microscopy (AFM).
Along with hexagonal patterns, which are indicative of a face-centered cubic (fcc) \{111\} plane (stacking disorder of the hexagonal planes, which has been a long-standing issue, e.g., as discussed in Refs.~\cite{Blaaderen1997,Zhu1997,Mori2011}, cannot be checked by surface observation alone), structures with dimensions of few tens of nanometers were observed.
Since gel-immobilized colloidal crystals were immersed in an aqueous dispersion of gold nanoparticles (AuNPs), these objects could be recognized as AuNPs attached on the surface.
Since AFM is not reliable for detecting edges of objects, the object size could not be very accurately determined.
By taking this limitation into account, the size of these objects was concluded to be the same as that of the AuNPs.
However, the possibility that the objects were AuNP-related ones but not AuNPs themselves cannot be completely ruled out.
The surfaces were deformed during the drying process.
Figure~\ref{fig:SEM_pre} is a scanning electron microscopy (SEM) image of a surface of a sample prepared in a previous study \cite{Kawakami2016}.
SEM observation were performed for samples after solvent substitution by an ionic liquid, as done by Kanai \textit{et~al.} \cite{Kanai2011}.
Although in AFM observations \cite{Kawakami2016}, objects that we recognized as AuNPs were often located at the basins at centers of triangles made of colloidal spheres on a hexagonal plane, in Fig.~\ref{fig:SEM_pre}, AuNPs (we recognized these from their sizes) were also observed on the periphery of colloidal spheres.
Some AuNPs may be detached in the drying process.
Solvent substitution by an ionic liquid provides an advantage:
The non-volatility of the solvent results in stability of materials against degradation due to evaporation of the solvent.
In this paper, we present a successive report on the hybrid plasmonic-photonic crystals with solvent substitution.

\section{Materials and Method}
\label{method}
To guarantee monodisperse colloidal particles, unlike in our previous study Ref.~\cite{Kawakami2016}, we used commercially available polystyrene (PSt) colloidal suspensions in this study.
PSt colloidal suspensions (Thermo Scientific) with 160 nm and 120 nm particle diameters were purchased. The volume fractions were $\phi$ = 10\% for both. Particle size polydispersity corresponded to 4\% and 5\% in terms of the coefficient of variation, respectively.
We report results using the 120 nm PSt suspension in this paper because in this study, the reflection peak wavelength could be successfully tuned to coincide with the LSP resonant wavelength of the AuNPs when the 120 nm PSt suspension was used.

The procedure for gel-immobilization of colloidal crystals was essentially the same as that by Toyotama \textit{et~al.} \cite{Toyotama2005}.
The gel regent ($N$-(hydroxymethyl)acrylamide, N-MAM), the initiator (2,2'-azobis[2-methyl-$N$-(2-hydroxyethyl)propionamide], VA), and the crosslinker ($N$,$N$'-methylenebisacrylamide, BIS) were purchased from Wako Chemical (081-01535, 929-11852, and 929-41412, respectively).

These chemicals were dissolved in Milli-Q water to prepare aqueous solutions (aqs.) with the same concentrations as in Ref. \cite{Kawakami2016}.
Then, these ($V_{N-MAM} = $170 $\mu$l of N-MAM aq., $V_{VA}$ = 100 $\mu$l of VA aq., and $V_{BIS}$ = 50 $\mu$l of BIS aq.) were mixed with the colloidal suspension ($V_{col}$ = 680 $\mu$l) after deionization by an ion-exchange resin (BidRad, AG501-X8).
Iridescence was confirmed for the deionized colloidal suspension before mixing and for the mixed suspension.
Reflection spectra of the colloidal crystals before and after gel-immobilization are shown in Fig.~\ref{fig:ref}(a).
The shift toward the long wavelength region for the gel-immobilized colloidal crystal is due to dilution resulting from the addition of gel materials to the colloidal suspension.
The lattice spacings between \{111\} planes assuming fcc structure are, respectively, $d_{111}$ = 190 nm ($\phi_{col}$ = 0.1) and = 220 nm ($\phi_{gel}$ = $V_{col}\phi_{col}/[V_{col}+V_{N-MAM}+V_{VA}+V_{BIS}]$ = 0.07).
Bragg wavelengths estimated based on these $d_{111}$ explain the shift in Fig.~\ref{fig:ref}(a). Also, the decrease in reflection peak intensity and broadening of the peak width can be considered to be indicative of decreasing crystallinity with a decrease in the thermodynamic driving force of crystallization.
After bubbling with argon gas, the mixtures were gellated in quartz cells with 1 mm thickness by subjecting them to ultra-violet (UV) irradiation. 
The durations of bubbling and UV irradiation were 5 and 30 min, respectively, the same as in Ref.~\cite{Kawakami2016}.

The gel-immobilized colloidal crystals were picked out from the cell and then immersed into the AuNP aqueous dispersion.
Several immersion durations up to 2 h have carefully been tested because a preliminary study \cite{Kawakami2014} suggested that an Au thin film formed after immersion for a long duration such as 6 h.
Among the tested durations, 2 h was specifically considered.
For a shorter duration, the number of AuNPs on the surface was low so that a large number of observations were needed because a lot of observed areas including few AuNPs.
On the other hand, for a longer duration, complex aggregates of AuNPs were formed.
The AuNP dispersion was synthesized according to the procedure in Morand \textit{et al.}'s paper \cite{Morandi2007}.
The averages diameter of the AuNPs was $\sim$40 nm. Then, the samples were immersed in an ionic liquid mixture for three days for solvent substitution.
The ionic liquid mixtures were composed of hydrophilic and hydrophobic ionic liquids.
The former was 1,3-diallylimidazolium bromide and the latter was 1,3-diallylimidazolium bis(trifluoromethanesulfonyl)imide, both of which were purchased from Kanto Chemical.
Kanai \textit{et~al.} \cite{Kanai2011} showed that the reflection peak could be tuned over a range from the middle of the visible light region to near-infrared region by varying the composition of the ionic liquid mixture. 
Reflection spectra were measured by a UV-vis-IR spectrometer (JASCO, V-670), and SEM images were obtained using a field-emission-type scanning electron microscope (Hitachi S4700).

\section{Results and Discussion}
\label{result}

To determine the composition of the ionic liquid mixture, swelling ratios of colloidal crystals without immersion in the AuNP dispersion were compared with the swelling ratio of the composition corresponding to $x$.
As in Ref.~\cite{Kanai2011}, we define $x$ as the volume fraction of the hydrophilic ionic liquid calculated using the volumes of ionic liquids before mixing.
The swelling ratios at each $x$ were calculated by comparing pre- and post-substitution photographs [Fig.~\ref{fig:swelling}(a)].
While the sample swells by substitution of the solvent by the hydrophilic one ($x$ = 1), it shrinks by substitution by the ionic liquid mixtures and the hydrophobic one. The photonic band structure (at least, its location) is, therefore, tunable thorough the $x$-dependent lattice constant variation of colloidal crystals.

Reflection spectra were measured for the gel-immobilized colloidal crystals without immersion in the AuNP dispersion. Peak wavelengths of reflection spectra at $x$ = 0.4 and 0.6 are plotted in Fig.~\ref{fig:swelling}(b). The dashed line is the peak wavelength of the sample without solvent substitution. Blue shifts are confirmed. Correspondence between shrinkage and blue shift was already reported in Ref.~\cite{Kanai2011}.
We find that the LSP resonant wavelength of an AuNP of 40 nm, $\sim$540 nm (\textit{e.g.}, Ref.~\cite{Yguerabide1998}), corresponds to $x$ slightly below 0.6.

\begin{figure}[htb]
\centering
\epsfxsize=0.2\textwidth
\epsfbox[85 10 200 530]{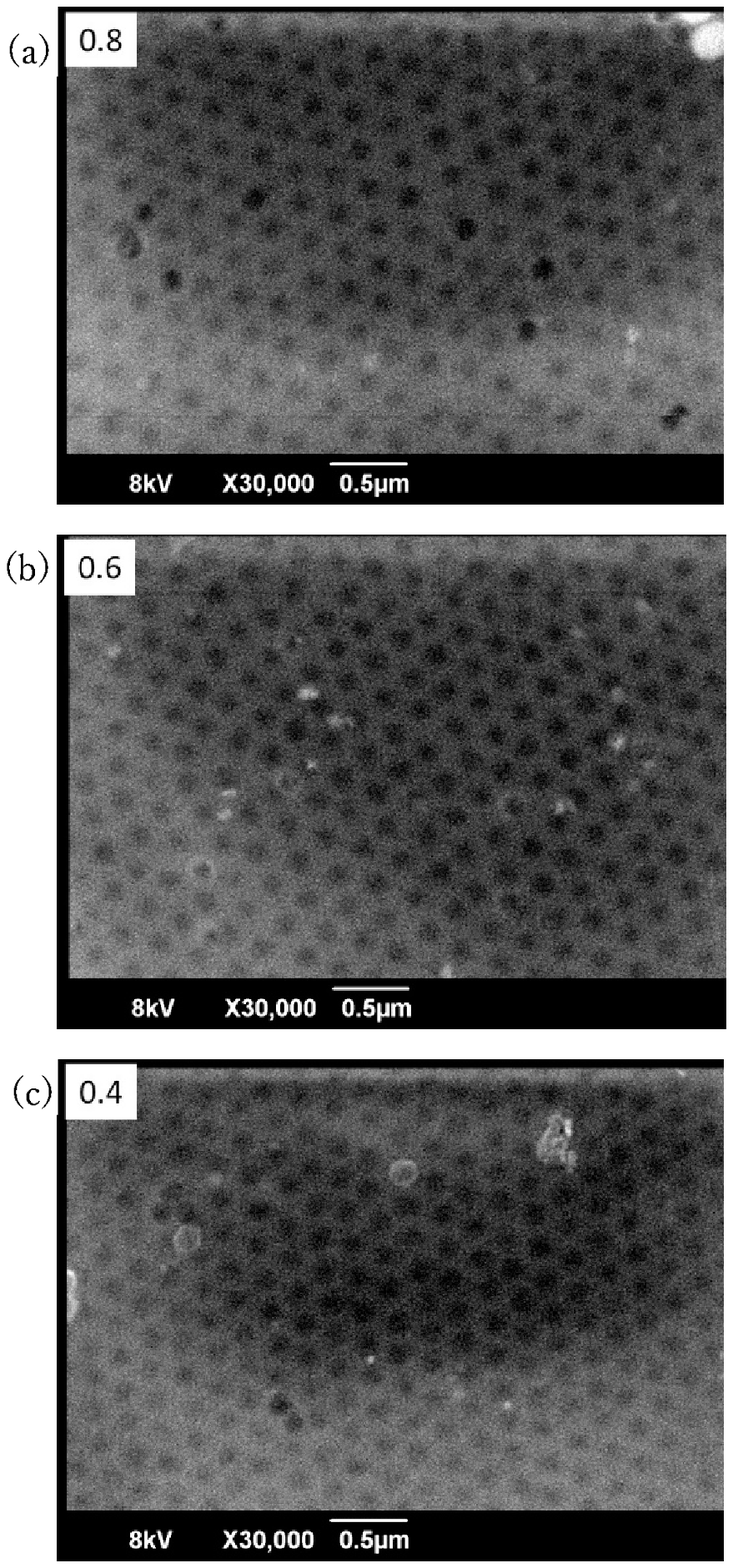}
\caption{\label{fig:SEM} SEM image of surfaces of gel-immobilized colloidal crystals after solvent substitution. ^^ ^^ Composition" $x$ is indicated in the top left corner of the panel. Lattice spacing reduces as $x$ decreases.}
\end{figure}

SEM images of surfaces of the gel-immobilized colloidal crystals after solvent substitution are shown in Fig.~\ref{fig:SEM}.
Hexagonal patterns observed by AFM, which are indicative of fcc \{111\} planes, as reported in Ref.~\cite{Kawakami2016}, have been confirmed. Lattice spacing decreases with a decrease in $x$ through the shrinkage of sample.
Further, some SEM images like those in Fig.~\ref{fig:SEM} suggest that AuNPs are attached on the surface, and some other SEM images indicate that treatments even in wet processes affect the surface structure. We consider that some AuNPs detached in the drying process, so careful treatment in wet process should be necessary.

Reflection spectra for AuNP-attached samples without and with solvent substitution by the $x$ = 0.6 ionic liquid mixture are shown in Fig.~\ref{fig:ref}(b). As expected, a peak shift was accomplished.
It should be noted that the reflection peak intensities significantly decrease by gel-immobilization and slightly reduce by AuNP adsorption and solvent substitution.
As already mentioned, however, the former can be attributed to the decrease in the volume fraction of the colloidal suspensions.
In future studies, we aim to separately determine the effects of this dilution and those of gel materials, which act as impurities inhibiting colloidal crystallization.
The absorption of light by AuNPs due to the LSP resonance is followed by emission of enhanced electric fields, which has strong directionality.
If one measures the light in a direction different from that corresponding to the enhancement, mere lowering of the peak intensities will be detected.
Further, if the photonic band structure varies, directional emission will be affected.
In addition, one should pay attention to the change in contrast between the dispersing particles and the dispersion medium during solvent substitution because the sharpness of the photonic band is affected by this contrast.

\section{Conclusion}
\label{conclusion}
We have tried to prepare hybrid plasmonic-photonic crystals on gel-immobilized colloidal crystals.
The solvent of the hydrogels has been replaced with ionic liquid mixtures to coincide the photonic band wavelength with the LSP wavelength of AuNPs attached on the surface of gel-immobilized colloidal crystals.
It has been successfully demonstrated that by varying the composition of ionic liquid mixture, the photonic band structure can be matched to the LSP resonance.
Significant electric field enhancement by hybrid plasmonic-photonic crystals on dried colloidal crystals is expected through photonic band tuning.

\section*{Acknowledgment}
\label{acknowledgment}
It is gratefully acknowledged that Prof. S. Hashimoto (Tokushima Univ.) provided gold nanoparticle dispersions.
This work was supported by a research grant from The Mazda Foundation in 2013.





\begin{figure}[htb] 
\centering
\epsfxsize=0.4\textwidth
\caption{\label{fig:SEM_pre} SEM image of surface of hybrid plasmonic-photonic crystal prepared by us \cite{Kawakami2016}.}
\end{figure}

\begin{figure}[htb] 
\centering
\epsfxsize=0.7\textwidth
\caption{\label{fig:swelling} (a) Swelling ratios corresponding to various compositions of the ionic liquid mixture.
(b) Reflection peak wavelength of gel-immobilized colloidal crystal after solvent substitution.
Shift toward a shorter wavelength after solvent substitution (dashed line) is seen in (b).}
\end{figure}

\begin{figure}[htb] 
\centering
\epsfxsize=0.4\textwidth
\caption{\label{fig:ref} (a) Reflection spectra of a deionized colloidal suspension (upper curve) and a gel-immobilized colloidal crystal (lower curve).
(b) Reflection spectra for hybrid plasmonic-photonic crystals without and with solvent substitution. The curve with the higher peak is for a sample without solvent substitution and the other is for one with solvent substitution. In both figures, the errors in reflectance were evaluated for several wavelengths by following time evolution. Note that the scales of vertical axes are different from each other.}
\end{figure}

\end{document}